\documentclass[aps,prd,preprint,superscriptaddress,showpacs,showkeys]{revtex4}

\usepackage{amssymb,amsmath}
\usepackage{graphicx} 
\usepackage{dcolumn}  
\usepackage{epsfig}   
\usepackage{pstricks}
\usepackage{ifpdf}
\usepackage{float}
\begin{document}


\title{Elementary Particles: What are they? Substances, elements and primary matter}


%
%

\author{D-M. Cabaret}
\affiliation{Couvent Saint Etienne, Jerusalem}
\email[]{dominiquemarie.cabaret@gmail.com}

\author{T. Grandou}
\affiliation{Universit\'{e} de Nice-Sophia Antipolis,\\ Institut Non
Lin\'{e}aire de Nice, UMR CNRS 7010; 1361 routes des Lucioles, 06560
Valbonne, France}
\email[]{thierry.grandou@inphyni.cnrs.fr}

\author{G.-M. Grange}
\affiliation{Couvent des Dominicains\\
1, Impasse Lacordaire\\
Toulouse\\
31400 France}
\email[]{gm.grange@gmail.com}

\author{E. Perrier}
\affiliation{Couvent des Dominicain{s}\\
1, Impasse Lacordaire\\
Toulouse\\
31400 France}
\email[]{perrier@revuethomiste.fr}


\date{\today}

\begin{abstract} 
The most successful \textit{Standard Model} allows one to define the so-called \textit{Elementary Particles}. Now from another point of view, philosophical, how can we think of them? Which kind of a status can be attributed to Elementary Particles and their associated quantised fields? Beyond the unprecedented efficiency and reach of quantum field theories the current paper attempts at understanding the nature of what we talk about, the enigmatic reality of the quantum world.
\end{abstract}


\maketitle

\section{\label{SEC:1}Introduction}
In a previous paper, the status of the wave function of Quantum Mechanics ($QM$) has been thoroughly analysed, in particular in the light of the most controversial \emph{epistemological} versus \emph{ontological} aspects of the matter \cite{wf}. Thanks to a few philosophical principles, both aspects were shown to be closely related the one to the other and, as a by-product, several $QM$ famous paradoxes and \emph{aporiae} were revisited and elucidated. Taking seriously `what $QM$ tries to tell us \cite{Mermin}', and in a most obstinate fashion as one could be tempted to emphasise, one of the key-points of this first analysis is that what is potential, although not actual as such, is real nevertheless \cite {Younan}. 

 This non trivial pattern of act and potency will be extensively applied throughout the present paper as it can be consistently posited that `what is potential is real'; not in itself, but inasmuch as it is ordered to \emph{act}. Potentiality is definitely part of the metaphysical structure of reality \cite{Foret}. 
\par
As just stated above, quite a lot of reputed curious cases become understandable in this way. The `crazy behaviour' of quantum entities, as it was coined long ago by R.P. Feynman, becomes perfectly rational and the `quantum enigma' \cite{Wolfgang} recedes in a substantial manner.
\par
In retrospect, getting back to the case of classical physics it is rather easy to understand why such philosophical considerations were unnecessary \cite{C&Q}. This is because classical physics essentially deals with actual realities while quantum physics delves into the question of knowing what they are made of.

\par
Implicit to this last statement is the bet that quantum physics is somewhat \emph{terminal} in the sense that no substructure should ever be expected, that is, that there is no deeper layer of the physical reality which would make of quantum physics some resulting form, generated out of this assumed deeper level. As will be explained elsewhere in more details \cite{C&Q}, there exist a number of good enough reasons to support this point of view, at both experimental and theoretical levels ${}^{\footnotemark[1]}$.
\footnotetext[1]{ `We don't have a physically plausible set of principles from which to derive quantum theory' \cite{Webb}.} 
  However it may be in the future, even if a deeper level of the physical reality was to be discovered, the pattern of understanding we introduce here would only be transposed to this new case.
\par\medskip
The $QM$ analyses are not terminal, though. $QM$ accounts for the physics inherent to the existence of a universal constant such as $\hbar$, but does it irrespectively of another fundamental constant which is $c$, the speed limit  ${}^{\footnotemark[2]}$
\footnotetext[2]{$c$ corresponds to the velocity of light in the vacuum. But what matters is that there exists, or not, a speed limit in the physical Universe, whatever its value and the physical process it is related to.} of any transfer of information or energy in the Universe. However, not only physics is ruled by both $\hbar$ and $c$, but the two theories, quantum and relativistic, call for each other as has been argued elsewhere \cite{QnatureofLinv}. This is why relativistic/covariant extensions of $QM$ have quickly been looked for. After some attempts, not conclusive enough, it was realised that another essential ingredient was in order, the concept of \emph{field} which is presented in the next Section and whose importance cannot be overemphasized. Quantum Field Theories, or $QFT$s for short, offer the most achieved theoretical realisations integrating the above three fundamental items, $\hbar$, $c$ and the notion of field, and their efficiency in microphysics is unprecedented. 
\par
It is therefore in the context of $QFT$s that \emph{Elementary Particles} are consistently identified as such, predicted and subsequently observed in high energy scattering processes. If a full unification of $QFT$s to Gravitation is not available yet (still, linearly approximated \textit{General Relativity} can be derived out of $QFT$ approaches \cite{Milton}), the former nevertheless are able to account for the observed Elementary Particles and their interactions. The highest energy resolutions obtained at $CERN$ in the $LHC$ experimental runs haven't disclosed Elementary Particles other than those already registered. This may have driven some expectations to despair \cite{Altarelli}, while others were confirmed in their working hypothesis that the elementary particle scheme offered by the so-called \emph{Standard Model of Particle Physics} (see below) has good enough chances to be effectively \emph{terminal} or at least extendable to much higher energy ranges \cite{Altarelli}.
\par\medskip
It is in this $QFT$ context that the topic of Elementary Particles will be coped with in the current article. Next Section will proceed with their proper definition as furnished by the above alluded Standard Model of Particle Physics. As defined in physics, Elementary Particles are compared to the ordinary realities of the physical world that philosophy qualifies as substances in Section III. To proceed further, we will see in Section IV that the notion of compositeness into whole and parts finds its limits with Elementary Particles, and has to step aside, leaving it to the notion of compositeness into matter and form. Accordingly, the central notion of a \textit{quantised field} is evaluated from a metaphysical point of view in Section V, and a further proposition concerning a would be primary matter, the \textit{materia prima} of philosophy, is exposed in Section VI. A conclusion summarizes our analysis.

\section{\label{SEC:2}What is an elementary particle ?}
To begin with, it matters to define what is normally understood by the notion of an elementary particle. In classical physics a notion whose role is paramount is that of the material point. A material point is a geometric point, devoid of any spatial extension and endowed with a certain mass, or quantity of matter. Of course, it is a pure mathematical idealisation, which is never exactly realised in physics, but whose convenience and fertility in classical physics provide it with the status of a true paradigm. The common experience furnishes many familiar examples, from the grains of sand to the distant stars which, at night, appear to us in the darkness of Heaven as tiny points of light. 
\par
This classic representation goes into the world of microphysics in which an electron, even if very enlarged, still appears as a small point mass. But in microphysics, the limits of the classical paradigm appear very quickly too. In fact, the world of very small scales, governed by the quantum laws, does not allow to maintain this representation of the corpuscles for a long time. One must substitute another concept which is that of a field, and more precisely, of a field associated to a given particle.
\par
Historically, the concept of field appeared in the last third of the 19th century, in the classical context of Maxwell's equations. But already, the difficulty is that this concept is much less intuitive than that of a material point and, what is more, appears as a quirk in the reductionist scientific context \cite{JMLL}: \emph{Already at the classical level, the notion of field requires the conception of a new mode of substantial existence, since it is necessary to give up the idea of the necessary and natural first qualities of matter in the sense of Descartes and Locke: A field has neither extension, nor solidity, nor figure, nor mobility to take up the Lockean list of these qualities} \cite{JMLL}.
\par
The concept of field becomes even less intuitive when, a few decades later, this same field acquires a quantum status, and receives the formulation of \textit{an operator valued distribution on a Fock space of states of the system being considered}.
However, as F. Wilcek invites us, it is worth noting that this concept of field, whether in classical or quantum form, has not yet shown a limit as to its applicability, whereas the Universe is observed over a substantial range of `sizes'.
\par\medskip
As a genuine paradigm, the field concept is powerfully synthetic of many aspects of physics at all scales. As for the corpuscular aspect one had started from in a first appraisal, it is recovered also as a limiting case, through a so called $LSZ$ process of reduction \cite{IZ}, operating within the mathematical framework of quantum field theories. In fact, as theorists often use to say to emphasize a relationship that is not obvious, this `deduction' of the corpuscular aspect is \emph{highly non-trivial}, since a whole body of theory is necessary for the interpretation of a $QFT$ in terms of Elementary Particles. This is the so-called \textit{Renormalisation Theory} \cite{IZ} whose last formulations, very algebraic, have only been proposed in recent years \cite{ConnesKreimer}.\par
Note that if all this remains quite complex technically, at the level of principles at least, these mathematical operations amount basically to the transposition of the famous `Born Rule' of standard Quantum Mechanics to the case of covariant theories of quantum fields.

Whatever the technical difficulties, and even at the cost of such elaborate algorithms, physicists have good reasons to maintain an interpretation of $QFTs$ in terms of Elementary Particles, because the experiments they perform, for example in accelerators, lend themselves naturally to `this form of reading or decoding' the results. The recent discovery of the famous Higgs boson at the $CERN$ $LHC$ in Geneva, has just shown this again.
\par\medskip
Experimental manifestations likely to be interpreted in terms of Elementary Particles are, in the jargon of theorists, \textit{point-events} or \textit{worldline's elements} of Minkowski's space-time, which, to be registered as such, require some localisation procedure in space-time. However, such localisation procedures are essentially arbitrary, being related to given experimental conditions. Now, the physical reality of the objects studied, the corpuscles, must not depend on the arbitrariness of the space-time localisations with respect to which it is apprehended. In other words, the characteristics which are intrinsic to the corpuscles must be indifferent to these particular localisations; and thus be, mathematically, invariants of the Lorentz-Poincar\'e symmetry group, which is a group of continuous transformations of point-event localisations in the \emph{affine} Minkowski space-time, preserving the causality attached to the existence of a finite speed limit \cite{tg1}.
\par\medskip
This is how we owe to E. P. Wigner a classification/identification of Elementary Particles as \emph{irreducible unitary representations of the Lorentz-Poincar\'e group}. Two invariant quantities characterise these representations. They are, on the representation considered, the values of the two  \emph{Casimir operators} of the Lorentz-Poincar\'e group, $P^2 $ and $S^2$, the squares of the energy-momentum operator, $P_\mu$, and of the spin operator, $S_\alpha$.

\par\medskip
In the definition of an Elementary Particle, this step is absolutely fundamental, but it is not sufficient, because according to this characterisation, the hydrogen atom in its ground state is also an Elementary Particle \cite{Barut}. We therefore see that the Wigner's classification must be supplemented with an additional criterion which is that of \emph{non-composition} in parts.  A particle, to be truly elementary, will be required not to be itself composed of sub-particles. 
\par
But how is checked the criterion of non-composition in a domain where one does not dispose of a hammer to break the particles in order to see whether they are made out of something smaller? 
\par\medskip
The observable Universe comprises four fundamental forces that are \emph{the electromagnetic force}, described by Quantum Electrodynamics ($QED$), \emph{the weak force}, which governs the decays of radioactive nuclei, \emph{the strong force} that governs the interactions of quarks and gluons constituents of the hadronic matter; and finally, \emph{the gravitational force}.
\par\medskip
The Standard Model, which unifies the first three forces in the same theoretical framework, is able to answer the question of knowing what is an Elementary Particle. 
\par
An Elementary Particle is a particle in the sense we have just defined, with the additional requirement that it can be considered as a point particle in all of its interactions with all other particles. This means that no internal structure can be ascribed to it: It is not composed in parts. Within some extra mild conventions \cite{MM}, a finite number of Elementary Particles can be identified in this way, which have become textbook material.

All Elementary Particles do not have the same life span, and, for example, the \emph{top} quark, last discovered, never lives but a tiny fraction of a second, of the order of $10^{- 5}$.
\par
The first three forces can be described in a unified way by the principle of \emph{local gauge invariance}, in the frame of what is called `the Standard Model of Elementary Particles'. The unifying principle underlying this model, the principle of {local gauge invariance} is the principle of a \emph{dynamical symmetry} determining the possible interactions between the various Elementary Particles, as represented in the few diagrammatical examples given below, FIG.1.
\begin{figure}[H]
\centering
   \includegraphics[width=\linewidth]{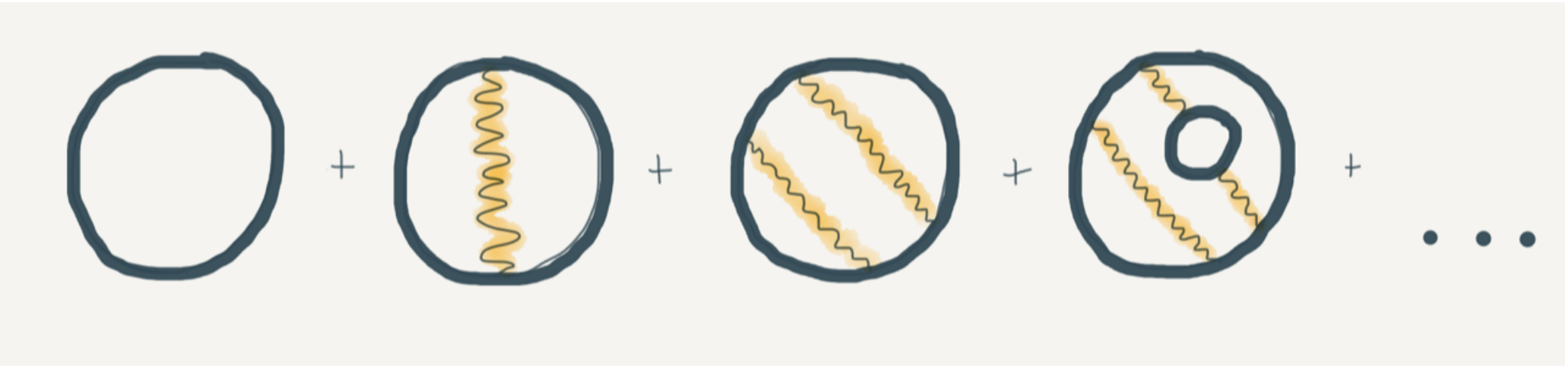}
   \caption{In $QED$, some `loop diagrams' associated to the logarithm of the vacuum transition amplitude $\langle0_+|0_-\rangle$ where $|0_\pm\rangle$ refer to the vacuum states at asymptotic times $\pm\infty$. Yellow wavy lines are photonic and plain lines represent the electrons/positrons.}
   \label{Fig.1}
\end{figure}
Here are, then, very summarised, the Elementary Particles as we describe them in contemporary physics. And given that knowledge, the question which is being addressed in this paper is this. 
\par 
{\textit{Are Elementary Particles realities equatable with well-known realities as cats, trees, or apples, all things that philosophy has long been naming substances? And, if they are not, in which way do they differ and how should we name them?}}
\par
What is at stake in this questioning is the attempt at understanding the nature of `what we talk about', that is, the nature of the very first elements of the material Universe.  Such an elucidation is crucial for any representation of the physical world, be it from the physical or philosophical point of view.

\section{Elementary Particles put to the test of \emph{substantiality}}
We have long been used to think and to speak of Elementary Particles as \emph{things} and there are reasons to do so. At the forefront of these reasons the fact that it is in this way that a link is obviously established between Elementary Particles and the World as we can apprehend it. Now this is not cogent enough a reason because this usage could be relevant of a thinking habit only, while the issue is rather to discern if the status of a thing can really be ascribed to Elementary Particles. On the one hand, experiments confront us to Elementary Particle behaviours which are those of substances, the `click' of a photon in a detector and measures thereof pointing to an \emph{individuality} (one click) endowed with a \emph{definite nature} (the capacity to activate a specific photon detector and not other kinds of detector) and to an entity being the \emph{subject of `a movement'} (as described by the photonic wave function), that is, undisputable characteristics of what makes of a something an authentic substance.
\par
On the other hand, taking a closer look at Elementary Particles it appears that their individuality is rather fuzzy as they fluctuate into \emph{quanta} of different natures and can also fade away whenever involved into some entangled states, all sorts of events definitely irrelevant to the classical world of substances. It is therefore appropriate to ask which kind of things are the Elementary Particles, or what is their \emph{thingness}.
\par\noindent
Let's specify further that a substance is a subject meeting the following characteristics \cite{Aristotle}.
\par\noindent

\par  (a) It is endowed with a certain actuality in that it exists in itself and does not belong to something else. 
\par (b) A substance is also supposed to possess the specific determinations of a given nature. 
\par
(c) A substance is the subject of a movement, which means it has a certain permanence through change.
\par
(d) A substance displays some individuality, with a certain conservation of itself (for instance a certain lifespan).
\par\medskip\noindent
Condition (a) is realised in the case of Elementary Particles, only under strict conditions, by means of  \emph{asymptotic} experimental protocols, contrarily to customary substances.\par\noindent
Condition (b) is realised only in that Elementary Particles inherit all of the determinations characterising the associated quantum fields, nature included, as will be discussed in details in Section V.
\par\noindent 
Condition (c) is not exactly met as quantum fluctuations make of the Elementary Particle-subject a somewhat unstable and blurred reality. \par\noindent
Eventually, condition (d) is hardly satisfied in general, not to speak of conservation that not all of Elementary Particles do enjoy, as the \emph{top} quark for example. As just evoked above, individuality is even more fundamentally questionable in the bosonic case (photons, gluons,~..) as well as in the fermionic case (electrons, neutrinos..) or in any situation of intrication. 
\par\medskip All of these limitations already suggest that Elementary Particles can hardly be considered as substances. While Elementary Particles display some behaviours which mimic those of plain substances, experimental and theoretical physics have made clear  however that their real status must be apprehended out of their very relation to the quantum fields physics has been led to associate to them. But if they are not substances, as trees, apples or cats are, then what are they?

\section{\label{SEC:4}Two kinds of compositeness and the issue of potentiality}

As evoked in Section II, the relation of Elementary Particles to their associated quantum fields is far from being trivial technically. It is this same relation which will be analysed now at a philosophical level, and to begin with, it matters to apprehend the very notion of \emph{compositeness} for the bodies composing the Universe,  as bodies appear to be either complex or simple.
\subsection{Two kinds of compositeness}
\par
In the common sense, we call complex, bodies which possess a certain unity while being made out of parts composing a whole. The fact that the whole may exhibit properties which the parts do not possess is what makes of the whole a unity in itself, not reducible to the sum of the parts, a case which is not unfamiliar to physicists \cite{Anderson}.  At our most general human level we are used to live among complex realities because, at the exception of light in some well defined conditions, we do not experience anything else. Even more so as we experience that some substances come to integrate new wholes, and become thereby parts of new substances. 

This implies that in order to understand complex substances, Science has proceeded, in a reductionist way, by taking composed substances into their simpler composing parts. It is in this way that these elementary structures have been brought to light and in all this, the composition of complex bodies according to the pattern of the whole and its constituent parts has been exemplified at lengths.
\par\medskip
This reductionist process however could not be expected to be continued endlessly. With the Elementary Particles the level of simple bodies is reached eventually, as they do not show a composition according to the pattern of `whole and parts'. By definition, there are no pre-existing parts out of which they would be composed. In this sense, and up to further possible discoveries, Elementary Particles deserve therefore to be called simple bodies. However, and no matter how simple they appear to be in regard to the `whole and parts' pattern, Elementary Particles are still composite from another point of view, which is that of a composition of \textit{matter} and \textit{form}.

\subsection{The inherent potentiality of simple corporeal bodies}
\par\medskip
The composition of matter and form has been extensively studied by many philosophers, although they never had the opportunity to check it in the realm of modern physics, as long put forth by several authors \cite{Tresmontant}. 
\subsubsection{On two aspects of the `form and matter' mode of composition}
In a nutshell, this mode of composition offers two interesting aspects. First, the distinction of  \emph{form} from  \emph{matter} and the identification of this matter. This latter aspect, evoked in this subsection, is dealt with in more details at the end of this paper. The former aspect is that form and matter enter into composition in a very specific way, which is that matter comes about as the \emph{principle of potentiality} and form as the \emph{principle of actuality} (note that because our knowledge relies on actual realities, we come to know bodies composed of matter and form thanks to the form; for instance we come to know photons through the fact that they have a polarisation-property and a frequency/wavelength/energy-property as well as a collective mode of behaviour). This built-in potentiality is deeply imprinted in the behaviour of the reality so composed of form and matter \cite {Foret, Aristotle2}.

In plain bodily substances are found both compositions (whole and parts; form and matter). For instance, Caesar's statue is a substance composed of Caesar's image (form) and marble (matter), which is itself composed from the crystal ordering (form) of carbonate minerals (matter), which are themselves composed from a particular ionised structure (form) of carbon and oxygen (matter).

When it comes to the quantum world, however, we cannot qualify Elementary Particles as substances in the sense we use for ordinary realities. In fact, what we have seen so far about Elementary Particles exhibits a specific composition between actuality-form and potentiality-matter. 
\par

Apart from the non-composition in whole and parts, Elementary Particles display stable and actual characteristics, that is to say: a) their invariant properties of mass and spin; b) they are describable within a synthetic representation space whose c) unitarity points to a peculiar persistent subject, and whose d) irreducibility indicates an elementary specific nature, distinct from any other. All these aspects account for the formal properties (which are definitely actual) of those realities we name Elementary Particles, and they would suggest that Elementary Particles are substances. However, we know that this conclusion would preclude the other aspects of Elementary Particles, relative to their matter-potentiality. 
\par
As we have seen, each of the characteristics of an ordinary substance has to be checked while, a) the Elementary Particle as such does not exist, formally, aside from its $LSZ$ projection in the corresponding renormalised $QFT$ framework ${}^{\footnotemark[3]}$,
\footnotetext[3]{ Where in full rigour the elementary particle is even defined as a limiting case: At asymptotic times and in the absence of interactions \cite{schweber, bigaj}.}and experimentally, aside from the protocol which identifies it as such, b) Elementary Particles are subject to quantum fluctuations, and c), their individuality is weak. Elementary Particles depart from plain substances in these regards, and do not raise to the level of actuality found in substances.
\par

In all, potentiality is so pregnant in Elementary Particles that they are not actual enough to be substances: In contradistinction to classical physics, which copes with a most superficial aspect of potentiality, that of movements, quantum physics is confronted to the potentiality that `quantum mobiles' display \emph{in themselves}. Indeed, were it not for the fact that quanticity is manifest also at \emph{mesoscopic} larger scales, Elementary Particles could be qualified as \emph{pre-substantial}. Be it as it may, it is the all remarkable that $QFT$s are able to cope with this unique mode of reality in a most efficient way \cite{Jackiw}.

\subsubsection{No physical reality prior to the elementary quantum fields}

\par\medskip
For Elementary Particles, the matter they are made of is not itself the form of something else, some pre-existing thing, energy included as will be argued in Section VI. As advertised above and described in details in the next Section, Elementary Particles cannot be conceived apart from the quantum fields they are associated to. It is it, the associated quantum field, which is this simple reality/body not composing pre-existing parts into itself. In other words, the matter of a quantum field is not a separate, actual reality which would become the principle of the upper level of reality that the quantum field is, like the marble is in the instance of Caesar's statue. Therefore, the matter of a quantum field is the most fundamental case of what a matter is since it has no underlying composition. Furthermore, noting that this very matter is deprived of any sort of determination, it can effectively be common to all elementary quantum fields, as it should to be consistently thought of as a primary matter.\par

\par\medskip\noindent
Concerning the quantum world as known of us and taken to be somewhat terminal, a materia prima, if any, should comply with the following requirements. 
\par
- It is present in every elementary quantum field while not being itself a thing since it is deprived of any formal determination. 
\par
- It would be what the quantum fields `are made of', while having no separate existence outside the quantum fields themselves.
\par
 - As a pure potency, it is the matter common to \emph{all} of the recorded elementary quantum fields.
\par\medskip\noindent
 Being \emph{the} pure potency, now, the materia prima introduces in the quantum fields, the principle of a fundamental \emph{instability}. This instability does not result into a would-be `becoming' for the quantum fields themselves, but introduces the associated Elementary Particles into all of their possible `becomings' : \emph{Substantial-like} changes, through creation/annihilation into new particles, or \emph{accidental} changes concerning the states of movements of Elementary Particles taken, then, as \emph{subjects} of movements. Now, to the quantum fields themselves, `nothing happens' indeed, at the very least because \emph{vacuum fluctuations} have no spacetime anchorage ${}^{\footnotemark[4]}$.
\footnotetext[4]{ More precisely, vacuum fluctuations take place at every point of the spacetime manifold $\mathcal{M}$ in a translational invariant way, that is in a way making no reference to any particular set of points in $\mathcal{M}$. In other words vacuum fluctuations have no possible history in $\mathcal{M}$. Because the quantised field energies are not localised in spacetime, this extends also to excited states.} 

  \par
\par\medskip
 In view of the characteristics recorded above, it should appear clear already at this level that this materia prima, if any, can hardly be mistaken for the quantum fields themselves \cite {dEspagnat}. It is tempting instead to think of it in terms of the ground state of the full system of elementary quantum fields in interactions. Tempting as it may look at face value, this point of view, which hits upon difficulties at a technical level (in particular, energy is no longer \emph{additive} a quantity in case of \emph{derivative couplings}) will be revisited in Section VI and finally refuted to let room for a possible and consistent resolution of the longstanding `primary matter' enigma.

\section{Quantum fields and \emph{elements}}
The notion of an Elementary Particle is no doubt rooted in the powerful \emph{material point} classical paradigm evoked in Section II. Now it appeared quickly enough that Elementary Particles scattering processes could not be accounted for if the basic elements of scattering processes are implemented as Elementary Particles as such. Quantum fields associated, each, to a given kind of Elementary Particle, were precisely conceived in order to account for these processes where the number and the nature of particles are subject to recognised changes. That is, if quantum fields have been introduced starting from the more intuitive notion of an Elementary Particle, it is now from the more involved notion of a quantum field that the deeper aspect of an Elementary Particle is revealed ultimately, since quantum fields are standing at a more fundamental level of the physical reality and since Elementary Particles cannot be understood as substances.
  \par
  It matters therefore to strive to understand and characterise what a quantum field is. For this, as will be proposed in subsection B below, the classical notion of an \emph{element} appears to be pertinent enough to make contact with the philosophical tradition ${}^{\footnotemark[5]}$.
\footnotetext[5] {For an overview on the elements in the Antiquity, cf. Rosemary Wright, Cosmology in Antiquity, Routledge, 1995, c. 6.}To begin with, subsection A provides a comparison of quantum fields to the more familiar wave functions of standard $QM$. Beyond the neat differences concerning their mathematical status, in effect, it is important to put forth the metaphysical similarities encoded in the two notions.

\subsection{\label{}From wave functions to quantum fields}
By writing a quantised field, like, in the photonic case and in the so-called \emph{Coulomb gauge} \cite{cohen},
\begin{equation}\label{photon}
A_\mu(t,\vec{x})=\sum_{s=1,2}\int {d^3k\over (2\pi)^3\,2\omega(\vec{k})}\,\biggl[\varepsilon_\mu^{(s)}(\vec{k})a_s(\vec{k})e^{-ik_0 t+i\vec{k}\cdot\vec{x}}+{\varepsilon_\mu^{(s)}}^*(\vec{k})a^\dagger_s(\vec{k})e^{+ik_0 t-i\vec{k}\cdot\vec{x}}\biggr]
\end{equation}
one has by construction (expansion on the \emph{plane wave basis}),
\begin{equation}\label{fakemove}\left({\partial^2\over \partial t^2}-{\partial^2\over \partial x^2}-{\partial^2\over \partial y^2}-{\partial^2\over \partial z^2}\right)\,A_\mu(x)=0\,.\end{equation}with the standard notation of $x\equiv (t,\vec{x})$. At first sight, this would suggest that the quantised $A_\mu$ field propagates in spacetime. Besides being an operator valued distribution, though, (\ref{photon}) makes explicit that the quantised $A_\mu$ field reads essentially as a continuous linear superposition of states, mutually orthogonal two-by-two,
\begin{equation}\label{perp}
s,s'=1,2\,,\ \ \ \varepsilon^{(s)}(\vec{k})\cdot \varepsilon^{(s')}(\vec{k})=\delta^{ss'}\,,\ \ \ \int \mathrm{d}^4x\  e^{-ik\cdot x}\,e^{-ik'\cdot x}=\delta^{(4)}(k-k')\,.\end{equation}
For the quantised field this pins down a \emph{potential} character which is in this respect exactly the same as that displayed by $|\Psi>$, the standard  state vector of $QM$ \cite{wf}. The field of (\ref{photon}) is in no way actualised/realised/projected in spacetime and doesn't propagate contrarily to what (\ref{fakemove}) indicates as an equation of propagation.  
\par
 Ambiguity comes from the following fact. As discussed in \cite{wf}, in the $QM$ formalism, $\Psi(t,\vec{x})$ can consistently be interpreted as a projection/collapse of the purely potential wave function $|\Psi>$ to the space point spotted by $\vec{x}$ at instant time $t$, which one may write as,
\begin{equation}\label{proj}
\Psi(t,\vec{x})=<t,\vec{x}|\Psi>\end{equation}It should be noted that (\ref{proj}) is written with a certain abuse in the notations because there is no time-operator in $QM$ \cite{jmll}. That is, a `ket' like $|t,\vec{x}>$ should rather be written and understood as $|x>_t$, the localisation state $|x>$ at instant $t$, in consistency with the $QM$ formalism. 
\par
Be it as it may
, this is no longer the case in the $QFT$ context of (\ref{photon}). In this latter case in effect, such a relation as $A_\mu(t,\vec{x})=<t,\vec{x}|A_\mu>$ does not exist, where a would be purely potential quantised field $|A_\mu>$ would play the role played by the wave function $|\Psi>$ in $QM$. This time in effect, not only $t$, but also ${\vec{x}}$ have no corresponding quantum operators. As well known, discontinuities between formalisms come into play when passing from $QM$ to $QFT$ where the genuine pendant of $|\Psi>$ (and not of $\Psi(t,\vec{x})$) is the expression (\ref{photon}). In the $QFT$ context, ${\vec{x}}$ and $t$ only refer to the localisation of a given point of the spacetime manifold, $\mathcal{M}$, on which a system of quantised fields is postulated. Equation (\ref{photon}) should therefore be read as the equation of movement satisfied by each plane wave \emph{individually},
\begin{equation}\label{move}\left({\partial^2\over \partial t^2}-{\partial^2\over \partial x^2}-{\partial^2\over \partial y^2}-{\partial^2\over \partial z^2}\right)\,e^{\pm ik\cdot x}=0\end{equation}that is, satisfied by each of the actual and physical realisation $A_\mu(t,\vec{x})$ is capable of. In (\ref{move}) the $4$-dimensional Minkowski space scalar product notation is used, $k\cdot x=k^0x^0-\vec{k}\cdot\vec{x}$. Satisfied individually by each plane wave $e^{\pm ik\cdot x}$, and thus by a continuous linear superposition of them, as is (\ref{photon}), of course (\ref{move}) implies (\ref{fakemove}). But it is important to realise that this doesn't mean that the quantised $A_\mu(t,\vec{x})$ field itself propagates in spacetime as if it was the \emph{subject} of a movement.
\par
Like the wave function $|\Psi>$ of $QM$, the quantised $A_\mu(t,\vec{x})$-field of $QED$ refers to a real, still potential  entity, as is explicitly meant through the continuous linear superposition of plane wave modes of propagation which are mutually orthogonal to each others \cite{wf}.

\subsection{Quantum fields as \emph{Elements}}
\subsubsection{Considering the element in itself}
\par With respect to the associated quantum field, Elementary Particles are not endowed with anything new so long as formal properties are concerned. These formal properties are the aspect of actuality that can be found in quantum fields, and that Elementary Particles inherit from the quantum field they are associated to. This dependence on the associated field extends to the fact that they are deprived of any autonomous existence. Even their collective behaviours, aggregating or repelling each others, are comprised in the formal determinations of the corresponding fields, that is in the very way their associated fields are quantised (see Equations (\ref{bose}) and (\ref{fermi}) below).
\par

With respect to the associated quantum fields however, Elementary Particles represent a higher degree of actuality as they are measurable entities whereas their associated quantum fields \emph{are not}, a crucial difference.
\par
All those attributes of quantum fields lead us to revisit the classical notion of element even if it has to receive a completely new, enriched, understanding. Quantum fields meet the notion of element in that they are, at the same time, the most fundamental corporeal reality, which stands apart from all other corporeal realities (due to its simplicity and to the fact that it is defined merely by a) properties and b) an inherent radical potentiality to any actualisation of those properties), and which doesn’t behave as any other corporeal reality. In this respect, Elementary Particles have to be conceived as elementary, not in the sense that they would be the most fundamental realities, but in that they are a highly selected case sharing in the only elementary reality which is that of the associated quantum field: An Elementary Particle is elementary in that it belongs to its underlying quantised field, able of the very corpuscular manifestation recorded as Elementary Particle.
\par
That being said, if quantum fields are considered in themselves, as elements, the relation of the formal properties to the matter they compose with is according to the composition of act and potency evoked already. Now, if the quantum field element is a simple reality in the very sense of Section IV (there is no composition of whole and parts) and if it is composed of form and matter, bearing all the formal properties, then the matter it is made of is pure potency to any formal determination. It is thus a matter common to all fields, in a sense which will be discussed soon, in more details. This, by the way, allows to render understandable that fields are in a position so as to be able to act upon each others, by energy transfers ${}^{\footnotemark[6]}$
\footnotetext[6] {.. even in the vacuum state of energy zero: In the loop diagrams of FIG.1, internal lines are endowed with non-zero energies/momenta.}as they act upon a matter which is common to all of them.

\subsubsection{The field-element as an environment}

To the field itself, `nothing happens'. As we have seen in subsection V.A, a quantum field has no actual anchorage in spacetime and does not propagate. Now,  propagations take place \emph{in} the field which, then, is  like an environment \cite{AZee}. Considered in itself, however,                                                                                                                                                                                                                                                                                                                                                                                                                                                                                                                                                                                                                                                                                                                                                                                                                                                                                                                                                                                                                                                                                                                                                                                                                propagation is not a thing, it is a movement, that is, an act. This act corresponds to a certain selection of the field formal properties along with a given amount of energy, which by the way, can also be viewed as a measure of the intensity of this act. The formal properties are thus with respect to energy as is potency with respect to act. This allows to make sense of the fact that this energy must be quantified in the very sense of measurable. In effect, a quantised field endowed with energy being a physical/material reality \cite{wf, Georgi}, any act in the field is an actualisation of the field formal properties according to a certain \emph{definite} quantity of energy. In a field, actualisation consists in a movement immanent to the field itself, an excitation of the field which propagates in the field \cite{AZee} and introduces a quantification in this very sense of a measurable amount of energy. Note that it is this measurable amount of energy which is at the origin of the Elemantary Particles corpuscular aspect, but non exclusively: Energy in the fields can manifest itself in other manners \cite{wf, Georgi} ${}^{\footnotemark[7]}$.
\footnotetext[7] {In Ref.\cite{wf}, this is expressed in a convincing way by Equations (11)-(14).}

\subsubsection{Elements as the matter of corporeal substances}                                                                                                                                                                                                                                                                                                       

The peculiarity of the field-element nature just evoked accounts for the fact that,                                                                                                                                                                                                                                                        again, Elemantary Particles can not be viewed as substances. Indeed, all bodies are made out of field-elements, not out of a bit of field-element itself, but through the determination of a given amount of excitation/propagation energy in the field-element. 
\par
As a result, it is not a bit of the field-element which becomes the matter of a higher level of a corporeal reality like hadrons and atoms; neither is it, keeping in mind the very definition of an Elementary Particle, the associated Elemantary Particles which become the matter of those higher realities. But the field-element is the matter of which all higher realities are made of in that, the field-element being excited-actualised, an excitation propagates within it with a determined energy: It is not a bit of the field which becomes the matter of a substance, but `the excitation/propagation' with its determined and measurable energy and formal properties, which is assumed by the form of a higher degree of a corporeal reality. This is an important \emph{articulation} at both physical and philosophical points of view.
\par\medskip
An analogy can help catching the point. Considering a wave propagating in the sea, the `substantialisation' would not consist in that a droplet of sea would separate from the sea to become a distinct entity, in which case in effect, the field-element would be the matter of a substance \cite{Couderc}, but in that `the wave in itself', is ordered to be what something else will be made of. The so-called \emph{Schwinger mechanism} may illustrate the point in the framework of $QFT$s. In this latter case in effect a peculiar excitation in the photonic field can give rise to an actual/measurable \emph{lepton-antilepton} pair provided that the photonic excitation is an energetic enough virtual photon \cite{HerbPidi}, and out of which pair something else could possibly be realised. In this case, it is made clear enough that this `something else' is not made out of the photonic field-element itself but of the lepton-antilepton pair raised out of the excitation in the photonic field-element.

 \subsubsection{In a picture}  
 The full seizure of the field-element can therefore be achieved according to the four aspects depicted on the Figure below.
 \begin{figure}[H]
\centering
   \includegraphics[width=0.7\linewidth]{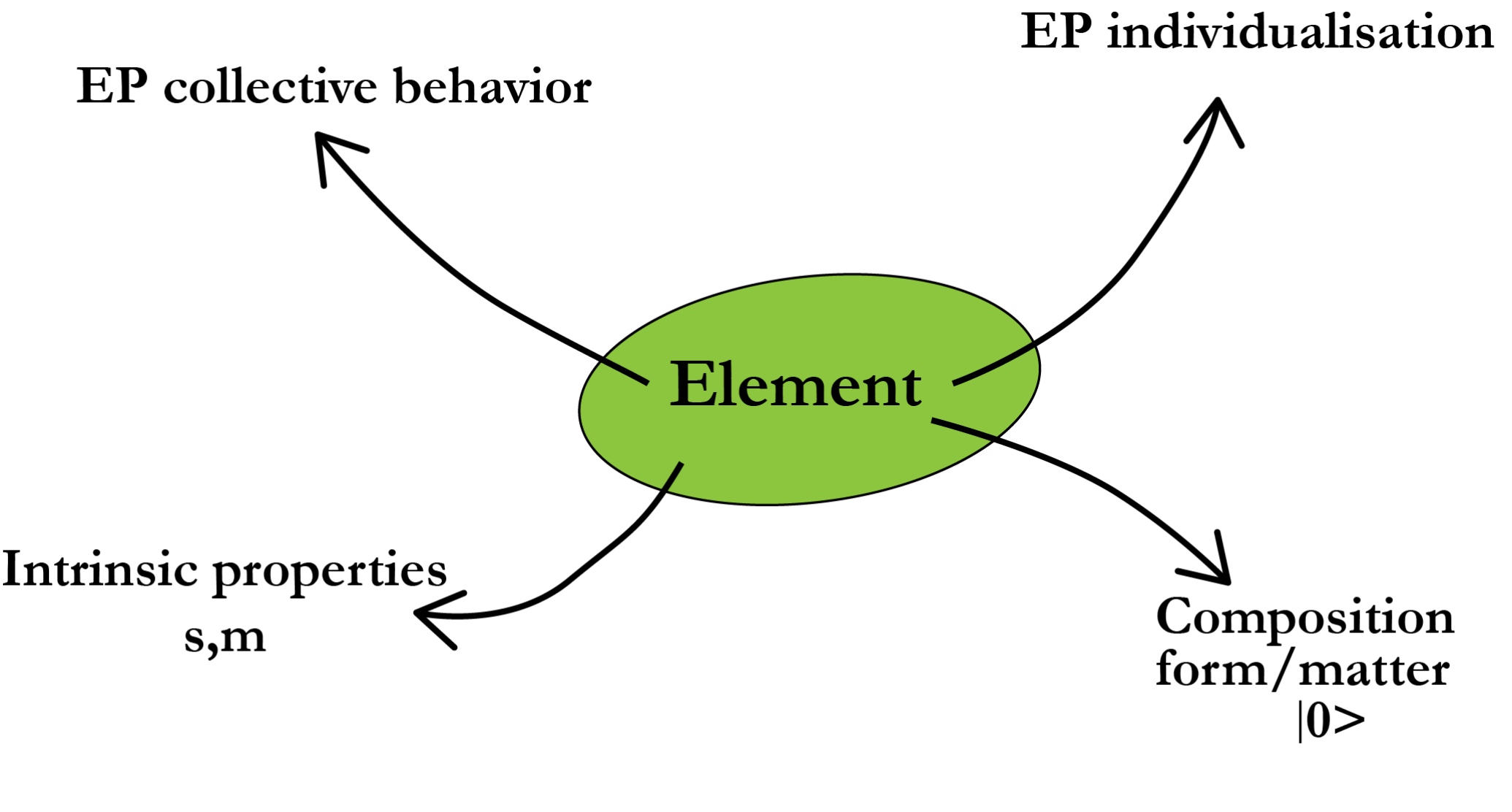}
   \caption{}
   \label{Fig.2}
\end{figure}

- The two dimensions indicated at the bottom of the Figure are relative to the fundamental identity of the quantised field understood as an element, that is an elementary composition of form and matter. The field-element is thus a material reality and even the very first of all material realities. Again, it is worth emphasizing at this place that the material thingness of quantum fields has received an indisputable support from the recent enough discovery of the \emph{Unparticle stuff} physics of quantum fields \cite{Georgi}, but is supported also by the much older Casimir effect (see Section VI.A). This, one may observe, is in contradistinction to the famous statement that `only correlations have a physical meaning, what they correlate do not' \cite{Mermin}, and to the most fashionable trend from there, which amounts to consider a Universe made of relations deprived of any physical substratum \cite{Tegmark}.
\par
-  On top-right of the picture, the (mild) individuation aspect is met at the level of the corresponding wave or corpuscular behaviors. This aspect is \emph{not} itself a formal determination of the field, which the particle would inherit from the field. Rather, the Elementary Particle is a field embodiment into some measurable amount giving rise to behaviors which are somewhat substantial-like in pre-determined experimental protocols. In its Elementary Particle aspect, the quantum field element is thus both an environment with respect to propagation/excitation processes, and it is what higher levels of the physical reality are made of (analogously, \emph{like} the matter of a substance; not forgiving the last point of paragraph V.B.3). In this way, the Elementary Particle can be seen the proper manner for the field-element to emerge from its full potentiality and to access actuality/measurability. Philosophically speaking, the Elementary Particle is not an \emph{accident} of the quantised field-element but its proper mode of actualisation in given experimental and theoretical highly selected contexts~${}^{\footnotemark[8]}$.
\footnotetext[8] {In the representation space of the quantised field's algebra where talking of Elemantary Particles makes sense, \textit{i.e.}, the Fock space (while an infinite number of other non unitarily equivalent irreducible representations are possible \cite{schweber}). This wouldn't necessarily hold true in another representation space, where finite energy physical manifestations in the fields may differ a lot. Quantised fields at high temperatures furnish a striking example \cite{Ingolf} where none of the experimental and theoretical conditions allowing for an Elementary Particle definition can be met. Physics phenomenology is encoded in the representation spaces.}

 \par \medskip

 It will be of interest (paragraph VI.C.1) to quote a complementary way to look at things, in which the Elementary Particle can be understood as borrowing its form to the corresponding field-element, and its matter to the available finite energy in the field. In this latter point of view one will state that the Elementary Particle individuates the form from the field-element into the definite energy lump the field possesses, and that it is, thereby, the field-element made measurable.
\par
 To sum up, it is in this way that the nature and deeper aspects of an Elementary Particle are revealed eventually, in a tight relation to the corresponding quantum field from which the Elementary Particle can never  be separated.  

\subsubsection{Our equations are more clever than we are (H. Hertz)} 

 Let us stress also that, at the exception of one of them only (see below), each of these points have their corresponding expressions within the $QFT$s mathematical formalism out of which they are being interpreted at a philosophical level. Recalling for instance the expression of the quantised photonic field (\ref{photon}) in the (\emph{physical}) Coulomb gauge,
 \begin{equation*}\label{photonic}
A_\mu(t,\vec{x})=\sum_{s=1,2}\int {d^3k\over (2\pi)^3\,2\omega(\vec{k})}\,\biggl[\varepsilon_\mu^{(s)}(\vec{k})a_s(\vec{k})e^{-ik_0 t+i\vec{k}\cdot\vec{x}}+{\varepsilon_\mu^{(s)}}^*(\vec{k})a^\dagger_s(\vec{k})e^{+ik_0 t-i\vec{k}\cdot\vec{x}}\biggr] \,,
\end{equation*}the terms $\varepsilon_\mu^{(s)}e^{-ik_0 t+i\vec{k}\cdot\vec{x}}$ refer explicitly to the formal properties that photons manifest as physical measurable entities, in terms of polarisation degrees of freedom and of frequency/wavelength/energy, while the quantisation rules obeyed by annihilation and creation operators, {\textit{i.e.}}, the \emph{commutation} relations
\begin{equation}\label{bose}
[a_s(\vec{k}),a^\dagger_s(\vec{k'})]= \delta _{ss'}\delta^{(3)}(\vec{k}-\vec{k'})\end{equation}encode the collective behavior of the photonic field quanta. As for the integral $\int d^3k$, it expresses the potentiality of the quantum field to be actualised in any of the possible values of k, each one providing the basis of a mild individuation, wave- or corpuscular-like. Of course the same considerations apply also to the free fermionic quantum fields projected out on plane waves, for which we have in $QED$, for the fermionic field operator and its \emph{conjugate} operator,
 \begin{equation}\label{fermionic1}
\Psi(t,\vec{x})=\sum_{s=1,2}\int {d^3p\over (2\pi)^3\,{\sqrt{2E_p}}}\,\biggl[ u^s(p)a_s(\vec{p})e^{-ip_0 t+i\vec{p}\cdot\vec{x}}+{v^s}(\vec{p})b_s^\dagger(\vec{p})e^{+ip_0 t-i\vec{p}\cdot\vec{x}}\biggr]
\end{equation}
 \begin{equation}\label{fermionic2}
{\bar{\Psi}}(t,\vec{x})=\sum_{s=1,2}\int {d^3p\over (2\pi)^3\,{\sqrt{2E_p}}}\,\biggl[ {\bar{v}}^s(p)b_s(\vec{p})e^{-ip_0 t+i\vec{p}\cdot\vec{x}}+{{\bar{u}}^s}(\vec{p})a_s^\dagger(\vec{p})e^{+ip_0 t-i\vec{p}\cdot\vec{x}}\biggr]
\end{equation}with for the quantisation, the \emph{anti-commutation} relations,
\begin{equation}\label{fermi}
\{a_s(\vec{p}),a^\dagger_{s'}(\vec{p'})\}=\{b_s(\vec{p}),b^\dagger_{s'}(\vec{p'})\}= \delta _{ss'}\delta^{(3)}(\vec{p}-\vec{p'})\end{equation}
In these expressions, $u^s$ and $v^s$ are four-component spinors, the analogues of $\varepsilon_\mu^{(s)}(\vec{k})$ in the photonic case, however more involved objects: To wit, with the plane wave 3-momentum $\vec{p}$ taken in the z-direction, one has \cite{MM},
\begin{equation}\label{ubar}
{\bar{u}}^s(p)=\left(\bigl[{\sqrt{E+p^3}}\frac{1-\sigma^3}{2}+{\sqrt{E-p^3}}\frac{1+\sigma^3}{2}\bigr]\zeta^s\,,\bigl[{\sqrt{E+p^3}}\frac{1+\sigma^3}{2}+{\sqrt{E-p^3}}\frac{1-\sigma^3}{2}\bigr]\zeta^s\right)\end{equation}where $\sigma^3$ is the third \emph{Pauli matrix} and the vectors $\zeta^s$ at $s=1,2$, label the 2 independent solutions $\zeta^1=(1,0)$ and $\zeta^2=(0,1)$. For a comparison with (\ref{photon}) , instead of (\ref{ubar}) and a similar expression for $v^s$, one has simply $\varepsilon_\mu^{(1)}=\varepsilon_i^{(1)}=(1,0,0)$ and $\varepsilon_\mu^{(2)}=\varepsilon_i^{(2)}=(0,1,0)$, within the same convention of a plane wave 3-momentum $\vec{k}$ taken in the z-direction. 
 \par \medskip
`At the exception of one point only', is a part of the sentence introducing this paragraph. This point has to do with the primary matter composing all of the elementary quantum fields and which, apparently, finds no corresponding term in quantum fields mathematical expressions such as (\ref{photon}), (\ref{fermionic1}), (\ref{fermionic2}), or any other similar expressions. This subtle issue is examined in the next Section where it should be made understandable that the situation cannot be different (see VI.C.2.).

\section{Quantum fields and the search for a primary matter}
What we have just seen so far confronts us with a striking fact: Quantum fields are a reality of this corporeal world, while, at the same time, their mathematical expressions display their inherent and total potentiality at its most (see Section V.A.). This duality gets translated into a capacity to receive (and be actualised by) measurable energy amounts. The relationship between potentiality and energy has been at the core of the speculations on what matter is at the primary level of corporeal reality. This research reveals to be extremely subtle indeed and deserves further considerations if the result is to comply with the absolute definition of a primary matter, philosophically conceived as an entity deprived of any form and any act in itself.
\par
In the context of modern physics, catching what this primary matter could possibly be identified to, is quite intricate a task. In the literature, several suggestions have been proposed, as for example, its identification to quantum fields themselves \cite{dEspagnat}. In the light of what we have understood so far, quantum fields cannot be thought of as a possible primary matter, in particular as they are not deprived of any formal determination (taken in itself, everything is potential in a quantum field except for its properties which, not only real, are also actual since they shape every excitation within the field). A sufficient refutation of this identification comes also from the following consideration on the \emph{Casimir energy}.

\subsection{Do Casimir effects reveal a primary energy?
}
Given an elementary quantum field-element, suppose that it would be possible to `peel out' the determinations brought about by its formal properties. In this way, one could think that the primary matter would then be reached as being the only thing left, purely potential and capable of `the whole quantity of the world'. It turns out that this naive idea, a sort of \emph{gedanken experiment} on the philosophical mode, is somewhat supported by a widespread comprehension of the infinite energy met in the \emph{Casimir effect}.
\par
The Casimir effect can certainly be viewed as one of the most fascinating and intriguing facts of $QFT$s and since its discovery by H.B.G. Casimir in 1948, it has been the matter of a lot of publications, reshuffled also by its possible links with the cosmological constant of \emph{General Relativity Theory}. In short the situation is this: Between two parallel perfectly conducting plates separated by a distance $d$, the Casimir effect is that of an attractive force between the plates which, per unit area, reads as

\begin{equation}\label{casimir}
{\mathcal{F}}=- \frac{\hbar c \pi^2}{240 d^4}\,.\end{equation}It is a purely quantum effect which has been experimentally verified to about $1\%$. The point is the striking simplicity of (\ref{casimir}). Up to the numerical factors in effect, (\ref{casimir}) could be derived on the bases of simple dimensional arguments \cite{IZ2}, and this is of course highly suggestive of the universal character of the vacuum quantum fluctuations that are taken to be at the origin of the Casimir effect.
\par
The force (\ref{casimir}) is obtained as the gradient with respect to the separation of the plates, $d$, of the \emph{zero point energy} of the vacuum fluctuations between the plates, ${\mathcal{E}}=+\frac{1}{2}\sum_n^\infty \hbar\omega_0(n)$, where the $\hbar\omega_0(n)$ are the eigenvalues of the free Hamiltonian of the system. At $3$ spatial dimensions this summation is divergent and can be regularised in different ways, each of them yielding the same result (\ref{casimir}) \cite{AZee}. For example, by imposing an upper cut-off $\Lambda$ one gets a result of order ${\mathcal{E}}\sim \Lambda^4$, that is an infinite amount of energy in the limit of $\Lambda\rightarrow \infty$. Borrowing to this infinite reservoir of energy it is speculated sometimes that an energetic enough quantum fluctuation has been able to generate our whole Universe \cite{Paris}.
\par

In the $QFT$ context, though, this `zero point energy' entails some arbitrariness as several choices of ordering for products of quantum field operators are possible, provided that they comply with the symmetries of the system. In the \emph{Fock space representation} of the \emph{fields quantisation algebrae} (\ref{bose}) and (\ref{fermi}), which is appropriate to the interpretation of $QFT$s in terms of Elemantary Particles (footnote [h]), the so-called `normal ordering' for products of quantum field operators can be chosen consistently and has the effect of turning ${\mathcal{E}}$ down to zero, while preserving Lorentz invariance \cite{Kleinert}.
\par
But there is more to it. As argued some years ago \cite{Jaffe}, the apparent universal character of (\ref{casimir}) is indeed an illusion `{\textit{that obscures the fact that the Casimir force originates in the forces between charged particles in the metal plates \cite{Jaffe}}}', and are relevant to Van der Waals forces between the plates (see also \cite{Kleinert}). 
\par
Such a statement, however, means that the force (\ref{casimir}) should be replaced by a null result in the limit of a vanishing $QED$ coupling constant, the famous fine structure constant $\alpha=e^2/4\pi\simeq 1/137$, which weighs the forces being exerted. 
\par The clue to this apparent contradiction is the fact that the idealised condition of perfectly conducting plates is accounted for by taking the opposite limit of $\alpha\rightarrow \infty$. It is in this limit that the coupling independent result (\ref{casimir}) is effectively recovered with its appearence of universality \cite{Jaffe}. It is worth emphasizing that it is precisely in the plainly physical context just evoked, relying on a one-loop $QED$ perturbative expansion that H.B.G Casimir himself derived his original result. Only in a second step, following a suggestion of N. Bohr, did he derive the same result within the much simpler idealised context which, since then, has become the standard one. It is in this way that the physical case has been traded for a much simpler but purely mathematical problem, with a poor relevance to physics \cite{Jaffe2}.
\par\medskip
In the end it is the whole family of Casimir-like effects which can be derived out of $S$-matrix calculations, that is out of Feynman diagrams with \emph{external legs}, without any reference to the vacuum and its fluctuations (corresponding to the Feynman diagrams of FIG.1 which have \emph{no} external legs) \cite{Jaffe}.  Accordingly, Casimir effects do not provide any physical support to the existence of a would be infinite reservoir of pure energy, the fluctuating vacuum state $|0\!>$, out of which quantum fields and their associated Elemantary Particles would be emergent entities. In this respect, much care must be taken in the interpretation of a fundamental $QFT$ expression like,
\begin{equation}\label{creation}
a^\dagger_{\vec{\varepsilon}(\vec{k})}|0>=|{\vec{\varepsilon}(\vec{k})}>\end{equation}
where the creation operator $a^\dagger_{\vec{\varepsilon}(\vec{k})}$ acts upon the vacuum state $|0>$ to produce the state of a given photon. The quantum theory cannot account for a creation process and even more so, disposes of sound arguments in favour of this radical impossibility \cite{jmll}. What is expressed in (12) is but the detailed balance of a \emph{change of state}, a fecund and well controlled notion of theoretical physics, while on the opposite, the notion of \emph{creation/anihilation} totally escapes its \emph{power of resolution}. This precious remark, never met in the scientific literature to our knowledge, is due to B. d'Espagnat \cite{dEspagnat,Grandou}.

\par
To sum up, Casimir effects do not really support the existence of a primary matter in the sense of a separate physical being, a would be reservoir of pure energy, deprived of any form, and out of which the first determinations of all of the material things would be made, elementary quantum fields to begin with ${}^{\footnotemark[9]}$.
\footnotetext[9]{Would it be so, and a reservoir of pure energy exist, then the `giver of forms' philosophical issue \cite{Ladriere} would come about immediately at this fundamental level in order to account for such a generation as that expressed in (\ref{creation}).} As a matter of fact, this rules out any identification of a primary matter to the quantum fields themselves, which are many and enjoy stable formal determinations.

\subsection{Are we done with a vacuum energy density?}
To our knowledge, physics wouldn't answer this question in a definite way. The issue of a non-zero vacuum energy is reshuffled by considerations bearing on the famous cosmological constant $\Lambda$, appearing in the master equation of the \emph{General Relativity} theory,
\begin{equation}
G_{\mu\nu}=8\pi G_N T_{\mu\nu}+\Lambda g_{\mu\nu}\end{equation}where $G_N$ denotes the Newton gravitation universal constant, $g_{\mu\nu}$ the spacetime metric tensor, $T_{\mu\nu}$ the energy-momentum tensor due to matter, and $G_{\mu\nu}$ the Einstein tensor. On very large scales, the Universe is admittedly described by a Friedman-Robertson-Walker metric tensor, \begin{equation}\label{FRW}g_{\mu\nu}=(1, -R^2,-R^2,-R^2)\,,\end{equation}
where $R=R(t)$, the so-called `scale factor', encodes the growth of the Universe. In this framework, it can be shown that a positive vacuum energy density $\rho_{vac}>0$ induces an acceleration of the Universe expansion while, on the contrary, matter induces a deceleration. A comparison with some cosmological data \cite{MM} would then indicate a vacuum energy density of about $\rho_{vac}\simeq (2\times 10^{-3} eV)^4$, and the question is therefore to know wether $QFT$s can address this question on account of the vacuum energy density \cite{Sakharov}.
\par
 $QFT$s must be \emph{renormalised}. They are deprived of any predictive power outside of this renormalisation procedure. In short, this is related to the fact that elementary quantum fields interact among themselves in such a way that there is no access to the \emph{bare} values of parameters like the masses and coupling constants entering the Lagrangian densities one starts from, like Equation (\ref{Lzero}) below. With more intuitive words, it is impossible to disentangle an electron from the cloud of quantum fluctuations surrounding it, so as to reach what would then be the \emph{genuine or bare} values of its mass and electric charge.
 \par 
 In the case under consideration, a whole series of diagrams like the ones depicted in FIG.1 in the case of $QED$, contribute to the total vacuum energy density. These diagrams display \emph{ultraviolet divergences} which, by supplying (\ref{Lzero}) with \emph{counter-terms}, can be cancelled term by term of a perturbative expansion in the bare coupling constant. Now, the leading divergence of the series in FIG.1 being quartic, the required cancellation amounts to an incredible fine-tuning phenomenon. As illustrated in \cite{MM} which is followed here, this exceptional tuning can be sketched on the bases of a simplified model, that of a scalar field endowed with a self-interaction $\lambda\varphi^4$, in a $4$-dimensional spacetime, with Lagrangian density,
\begin{equation}\label{Lzero}
\mathcal{L}=\frac{1}{2} (\partial\varphi_0)^2-\frac{1}{2}{m_0}^2 \varphi_0^2-\frac{\lambda_0}{4!} \varphi_0^4-\rho_0+\cdots \end{equation}where the subscript $0$ indicates bare quantities: Field, mass, coupling constant and vacuum energy density, $\rho_0$, all of them depending on a cut-off parameter $\Lambda_{cut}$ introduced to regularise logarithmic and power-like ultaviolet divergences. The renormalised vacuum energy density can be calculated out of (\ref{Lzero}) and reads as, \begin{equation}\label{rhos}\rho_{vac}=\rho_0(\Lambda_{cut})+c\Lambda^4_{cut}\,,\end{equation}where the last term is an example of a counter-term, and $c$ some constant. As there is no access to the energy density $\rho_0$ in (\ref{Lzero}), it is but a free parameter which is chosen so as to make the right hand side of (\ref{rhos}) coincide with the experimentally determined value of $\rho_{vac}$, the one quoted above on the bases of cosmological considerations.
\par
Now, most recent experimental runs have supported the relevance of $QFT$ descriptions of microphysics up to energy scales of some $10\,TeVs=10^{13}\,eV$s. 
Inserting this energy scale in (\ref{rhos}), one must therefore check a numerical match of
\begin{equation}(2\times\, 10^{-3}\, eV)^4=\rho_0(\Lambda_{cut})+c(10^{13}eV)^4\,.\end{equation}Though arbitrary, $\rho_0(\lambda_{cut})$ is therefore to be chosen so as to cancel out some $10^{52}eV^4$ and leave such a small amount as $10^{-12}eV^4$. This is extraordinary for a tuning phenomenon which does not meet any similar instance in the whole context of $QFT$s, and remains peculiar to the renormalised vacuum energy density calculation. Moreover, this milli-$eV$ scale does not seem to present any relation to the whole realm of particle physics. 
Such an astounding match renders dubious the assumption that quantum vacuum fluctuations could be responsible for the observed value of the cosmological constant. Furthermore one cannot exclude either that a non-zero vacuum energy density, if any, does not couple to gravity \cite{lehman} and that the cosmological constant may have an origin other than quantic. In any case, we do not dispose `of a natural theory of the cosmological constant' \cite{Altarelli}.
\par
To sum up, in view of the dubious aspects of these estimates, together with the speculative assumption at play ${}^{\footnotemark[10]}$,
\footnotetext[10]{Homogeneity and isotropy of the Universe, for example, can be viewed as a questionable hypothesis \cite{klnrt}} (\ref{FRW}) to begin with, one should not conclude out of this \emph{medley} of gravitation theory with $QFT$s, in favour of a non-zero vacuum energy density. No doubt that so long as a unified theory of gravitation with quantum physics is not available, this issue cannot be addressed with enough consistency and reach \cite{Peskin}.
\par\medskip
There would remain the issue of \emph{dark matter and dark energy} according to which a few percents only would be perceived of the whole Universe content, some 5\% maybe. It is worth recalling that dark matter was introduced in order to account for the galactic anomalies observed in some velocity distributions, and dark energy to account for the expansion of the universe. Up to now, in spite of several programs of research devoted to this issue, all over the world, no sign has ever been registered which would accredit the existence of a dark matter or energy, as an observed physical reality, and not a word can be said concerning their nature. They do not provide enough integration to the $QFT$ context to allow for an understanding in the framework we are dealing with. Concerning dark matter, moreover, it has been shown that if the proper mathematical status of the observational data set is taken into account, as a model independent projective geometry structure ${}^{\footnotemark[11]}$
\footnotetext[11]{Model independent as the only assumption at play is that the Universe admits a finite velocity limit to energy and information transfers (See footnote [2]).}then the problem at the origin of the dark matter hypothesis simply fades away \cite{Jaco}.

\par\medskip
Finally, there does not seem to be any compelling enough reason to suppose that a quantum non-zero vacuum energy density would suffuse the Universe, \emph{condensate}-based arguments included, as suggested in the Appendix. There is no quantum reservoir of a `pure energy' out of which corporeal realities would proceed in some unknown way, left open to the \emph{giver of forms} \cite{Ladriere}; and likewise, there is nothing like a `pure energy' in the sense of an energy which would be deprived of any sort of form. Energy comes about with some support, a something whose it is \emph{the} energy. This is a non-trivial point \cite{next}.
\par\medskip

A key-point of the current paper is therefore that quantum fields are the very first material realities composing the Universe, a claim whose philosophical consequences cannot be overemphasized. Quantum elementary fields are simple corporeal realities, elements, and do not result from any composition of a form with a pre-existing physical reality. Still, quantum fields are material, they have physical dimensions and exert measurable interactions on other material objects, like the conductive plates of a Casimir effect, for example. Whatever their given specie, `a matter' is necessarily common to all of the quantum elementary fields, though not existing separately, on its own.

\par\medskip
At this place, one cannot help recalling the amazing depth of Aquinas thoughts so many centuries ago \cite{Aquinatus},
 \par
 {\textit{ When we speak about the matter existing in this thing, we let aside the consideration of a matter understood absolutely. […] Therefore, the consideration of the matter of this thing is not the consideration of the matter understood absolutely, but of the matter existing under determinate dimensions. Therefore, what pertains to the matter qua absolute and primary does not necessarily pertain to the matter existing in this thing, as it is existing in this thing; because in this latter case, one drifts apart from the consideration of the primary matter. Therefore, the matter existing in this thing is not in potency to the whole quantity of the world.}}
 \par\medskip
  One may observe that this statement answers `the philosophical gedanken experiment' introducing Section VI.A. 
  \par\noindent
Specifically, the statement above bears on matter and a would-be primary matter and not on energy which we know by now that it is equivalent to matter. It is interesting to implement this equivalence into the words of Aquinas as, in this way, one reaches the conclusion that, again, energy can not be primary matter.

\subsection{\emph{Primary matter} is not energy}
If the quantum field is not made out of an energy which would play the role of a pre-existing matter, and if it cannot be identified to energy itself, what, in the end, makes of it a material reality, as physics attests it is? What is it that makes of it the most fundamental material reality of the Universe?
\par
Here it is necessary to get back to the composition of form and matter such as exposed in Section IV.B. Matter is what a thing is made of, the co-principle composing into a \emph{being}. It is in this sense that the field-element is the matter of the corporeal realities of the world. From this point of view, in a somewhat superficial way, the field element appears as a matter endowed with energy. However, it is not an energy deprived of form: It is always a peculiar energy to which are attached the field formal properties. As we have seen already, an energetic enough virtual photon decaying into a real lepton/anti-lepton pair illustrates the point as the virtual photon is not an energy as such, \emph{in abstracto}, but an energy in a given form, here photonic.
  \subsubsection{What elementary quantum fields are made of }
The quantum field element though must be considered in itself, in a more accurate manner and independently of its composition into higher degrees of the physical reality. And then, three fundamental and distinct aspects come about. 
\par
(i) The formal properties which allow to discriminate the various fields, 
\par
(ii) the excitation energy in the field, with regard to which it behaves as an environment,
\par
(iii)  the potentiality of the quantum field element.
\par
 It is this third aspect which constitutes the matter of the field in that it cannot be mistaken for either the formal properties (they are actual) or the excitation energy in the field (it is a process of actualisation). Potency is what the quantum field element is made of. However bizarre this statement may sound, this dimension of potentiality, inherent to quantum fields, verifies the notion of a matter here understood in a technical sense, both philosophically (what a thing is made of and in composition with the form) and physically (the quantum field differs from the classical field precisely in this potentiality which makes it fluctuate into other quantum fields).

It turns out that $QFT$s are able to characterise this materiality-potentiality in a definite manner through their consideration of the quantum field fluctuations. These quantum fluctuations in effect are able to manifest the distinction between the quantum field element taken as a material reality plagued with a fundamental potentiality, and the quantum field taken as a determined environment for the propagation of definite quantities of energy. In this latter case the quantum field appears filled up with energy since it is the potential subject of all of the excitations propagating through it. In the photonic case, these quantum fluctuations admit the diagrammatical representation of FIG.3.

\begin{figure}[H]
  \includegraphics[width=\linewidth]{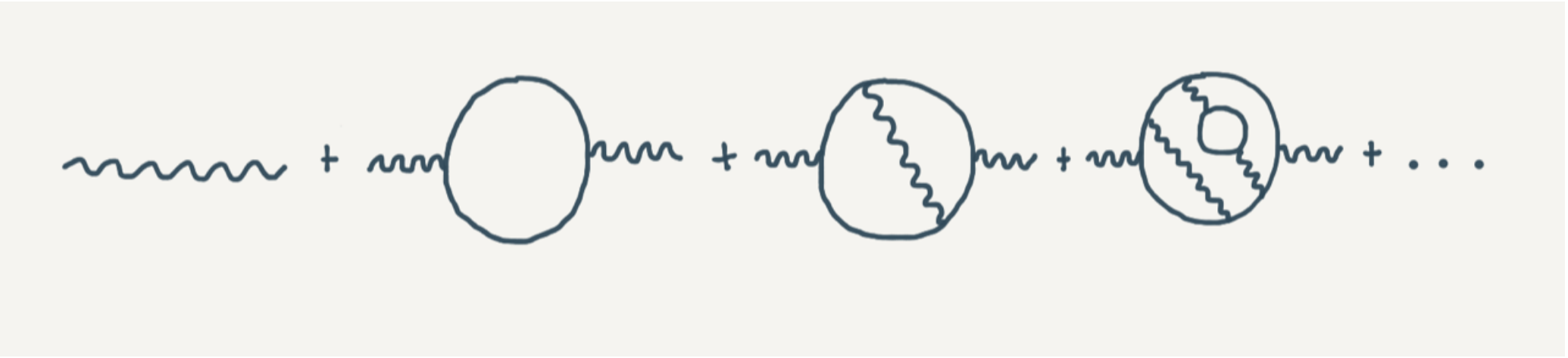}
   \caption{In $QED$, some diagrams representing the quantum fluctuations of a photonic state.}
   \label{Fig.3}
\end{figure}
\par\medskip

Now, taken as a material reality in its own right, the quantum field shows up as a pure potentiality to all of the excitations determined by its formal properties, together with its recorded interactions with the other quantum fields. In this sense in effect it is possible to posit that the matter of these excitations, all of them proceeding along with the fields formal properties, can be identified to the fields potentiality; or in other words, that the fields potentiality plays the role of the fields universal matter \emph{contracting} the various forms of the fluctuating quantised fields.
\par
Still, this has to do with excitations in the fields and the consideration above can therefore be taken as a point of view because, as quoted in Section V.B., another perspective amounts to look at the excitations as borrowing their forms to the fields and their matter to the energy in the fields. To avoid the ambiguity related to the possibility of these two different points of view it is appropriate to consider the fields bereft of any energetic excitations, that is fields in the vacuum state so defined by the equation  $P_\mu[0>=0$. 
\par
 Staring at the quantised field expressions it would seem that if there is no energy, no $k$ in (\ref{photon}) nor any $p$ in (\ref{fermionic1})  and (\ref{fermionic2}), then there is no longer any formal determination, and thus, \emph{stricto sensu}, nothing (\emph{no thing}) left at all.  However, if a classical field can be turned out to zero, an elementary quantised field cannot: Even endowed with a null energy a quantized field doesn't simply fade away, it fluctuates, a point of utmost importance as H.B. Casimir himself considered ${}^{\footnotemark[12]}$.
\footnotetext[12]{No contradiction with the quantised field expressions just evoked if one keeps in mind that they refer to continuous linear superpositions of actual/measurable realisations which, as such, have non vanishing energies even though the case of zero energies is formally included in the integration domain of (\ref{photon}), (\ref{fermionic1})  or (\ref{fermionic2}). In $QFT$ calculations and for massless fields, this results into so-called \emph{infrared and/or mass singularities.} While the latter cancel out when properly dealt with, both at zero and non-zero temperatures~\cite{tg}, in concrete physical processes (emission for example) the former may require the introduction of a resolution parameter $\Delta E$ precisely related to the experimental device ability to detect the tiniest energy amounts \cite{Peskin1}. This restores for the infinite set of potentialities of (\ref{photon}) the necessity of having a non zero energy in order to be actualised/measured.}   
 \par
  \subsubsection{Primary matter as vacuum-exhibited }
This means that, though deprived of energy, the vacuum state $[0>$ is not nothing. In loop or coupling constant \emph{perturbative} expansions, the probability amplitude of the vacuum state \emph{persistence} (\textit{i.e.}, its stability) is consistently identified to an infinite sum of fluctuations, the so-called vacuum diagrams such as those depicted in FIG.1 in the case of $QED$. An important issue is therefore to know wether these fluctuations are real, concrete phenomena, or nothing else than some formalism's \emph{artefact}, just operating in a successful algorithmic way.
 \par
  It turns out that most physicists would answer that these vacuum fluctuations are real, physical phenomena \cite{Weinberg}.  No more no less physical, at least, as are the quantum fluctuations in calculations of $S$-matrix elements \cite{Jaffe}. Now, that the latter are physical is agreed on by now as being definitely supported by a series of high precision tests confronting the $QFT$ theoretical predictions with impressive success \cite{Peskin2}. One may quote also the striking experimental confirmation of a running strong coupling constant in $QCD$, again the result of the quantum fluctuations \cite{Huang}.
   \par
    In view of so convincing experimental confirmations of the theoretical predictions one can feel entitled to think with A. Einstein that one possesses sound `elements of reality'; and so it is posited in the current paper.
    \par

In this way, the reality of a matter in the quantum fields can be seized independently of any energy considerations of energy in the fields, and it shows up as a pure potentiality to all of the possible definite energy actualisations the  quantised fields are capable of, given their mutual interactions. This state of affairs being common to all of the recorded elementary quantum fields, it is accordingly legitimate to state that the quantum fields vacuum state \emph{exhibits} `a matter' which is primary and accounts to `the whole quantity of the World' \cite{Aquinatus}; and that, though being really distinct from the quantum fields themselves, this primary matter does not enjoy an existence which would be separated from these elementary quantum fields: Elementary quantum fields would remain the first physical realities of the material Universe and their matter, primary, would compose with their forms as the fields co-principles.
\par
With these vacuum fluctuations it is worth pointing out that one is facing \emph{a reality totally deprived of actuality,} an exceptional fact in the whole realm of physics which $QFT$s only have been able to manifest  ${}^{\footnotemark[13]}$.
\footnotetext[13]{It must be realised that it is not so for quantum fields endowed with \emph{definite} amounts of energy: Though not necessarily localised in spacetime, a \emph{definite} amount of energy is no doubt an actual reality.}

\par\bigskip
In all, the characteristics of a primary matter such as exhibited by.the vacuum state of elementary quantum fields can be summarized as follows, completing the first attempt given in Section IV.B.2.
 
- It is pure potency, deprived of any formal properties, of any act (of being) and thus, not a \emph{thing}.
\par
- It is what elementary quantum fields are made of, while having no separate existence outside of the quantum fields themselves.
\par
- It is `the matter' common to all elementary quantum fields, allowing them to fluctuate the ones into the others according to their recognised interactions,
\par
- And last but not least, this primary matter is not an energy. This complies with the fact that energy doesn't show up free of formal determinations.
\par\bigskip
To sum up, the difficulty at understanding what primary matter can possibly be, is entirely related to its definition as \emph{absolute potency}, about which, as such, only negative statements can be formulated. This explains also why nothing in the quantised field mathematical expressions (\ref{photon}), (\ref{fermionic1}) and (\ref{fermionic2}) could be ascribed to the primary matter they are made of~${}^{\footnotemark[14]}$.
\footnotetext[14]{Maybe not that much of a limitation to H. Hertz sentence introducing V.B.5., than the fact that our equations keep on teaching us, be it in a negative way.}Still, primary matter is  a reality, but a reality deprived of any sort of actuality.  However striking it may be, primary matter does not come about as a free hypothesis but instead as a conclusion that a philosophical investigation of $QFT$s forces upon us.

  \par

\section{Summary}

\emph{The starting point of the metaphysical analysis is not in the texts, but in the reality itself in its full glory. Never before has the reality been known of us to such an extension, depth and reachness. Never before have the metaphysical problems questionned the human mind so clearly}, as C. Tresmontant claimed decades ago on the bases of the twentieth century's scientific developments \cite{Tresmontant}.
\par
It is a widespread conviction in effect that modern physics begs crucial philosophical questions in a way such that `Physics needs Philosophy, Philosophy needs Physics' \cite{Rovelli}. This very deep link between physics and philosophy is indeed historical, each discipline having motivated and conditioned the evolution and progresses of the other one.
\par
Through centuries, however, this reciprocal influence has fluctuated in intensity. While classical physics could allow itself to be somewhat sloppy, for reasons which one can easily understand by now, quantum physics confronts us to philosophical challenges which one can no longer avoid, unless Science, physics in the current case, is denied any possibility of accessing genuine elements of reality. This philosophical attitude exists also and can find perfect {`closed-form'} expressions \cite{Bachtold}, but is no doubt frustrative while History shows that humans have always striven to circumvent it, and  quite successfully indeed.
\par\medskip
Having reviewed what physics identify as Elemantary Particles it has appeared quickly enough that the singularity of Elemantary Particles is that they do not qualify to the perfection of being substances. By contrast, the ancient notion of substance is useful to understand how and why Elemantary Particles differ from common things offered to our eyes, and to make it clear that Elemantary Particles definitely belong to the quantum world. They are properly defined and understood within the context of renormalised $QFT$s, and experimentally checked through accurate protocols, within a background dependent \emph{metric space}, Minkowskian ${}^{\footnotemark[15]}$.
\footnotetext[15]{Extension to background independent spacetimes is quite involved but can be shown to preserve the reliability of the local Minkowskian definition, as it should \cite{Rovelli2}.} 
 \par
As a consequence, the deeper aspect of the matter turns out to be that of the quantum fields associated to the Elemantary Particles. As long encoded in renormalisation theory and the $LSZ$ reduction formulae, in effect, Elemantary Particles cannot be disentangled from the fields they correspond to (though not through a \emph{one-to-one} correspondance because the $LSZ$ procedure operates as a projection \cite{AZee}). But whereas Elemantary Particles are measurable manifestations of the quantum fields, quantum fields themselves, like the wave function of $QM$ totally escape measurability. \par
It is thus this very notion of a quantised field which deserves to be analyzed in more details, comparing it, to begin with, to the maybe more familiar notion of a wave function. Then the ancient notion of element is revisited to offer a deeper understanding of what a quantum field is.  This supposes the clarification of composition in material bodies. Composition, in effect, can be either complex or simple, proceeding along the modes of `whole and parts' and `matter and form'. The latter reveals to be the only mode of composition relevant to elementary quantum fields and this immediately begs the question of knowing which kind of a matter, then, is involved in this peculiar mode of composition. Positing that the so-called standard model of Elementary Particles is somewhat terminal, that no deeper layer of the physical reality should be expected, this primary matter would accordingly be nothing but the \emph{materia prima}  long conceived and defined by philosophers.
\par
 It is true that  attempts at identifying a primary matter out of quantum physics is not new. Proposals still, have never been convincing. Elementary quantum fields themselves cannot be considered as pertinent candidates, no more, quite surprisingly, than their zero modes, as argued in the current paper along with the issue of a vacuum energy density.  In the end, it turns out that quantum fluctuations are able to display the typical behaviour of primary matter, which is not a `thing' existing in itself, separately: Though not nothing, primary matter is not a thing, but a potency-for-a-corporeal-thing-to-be. 
 \par
 Quantum fields as elements are accordingly the very first realities composing in the physical bodies of our material Universe,  a claim whose philosophical and physical consequences are not innocuous:  Among them, the issue of the `giver of forms' can be understood as irrelevant to this fundamental level of the physical reality, while, throughout the present analysis, the singular status of energy and its role are implicitly singled out \cite{next}.
 \par
The present analysis extends to the relativistic quantum theory of fields what has been undertaken about the wave function of Quantum Mechanics \cite{wf} and, if pertinent,  it should allow us to acquire a better understanding of the nature of what we talk about: The `crazy' objects of the quantum world.

\section{appendix}
In $QFT$s a non trivial case is that of \emph{condensates}. More specifically, are condensates intrinsically related  to vacuum transition amplitudes, $\langle 0_+|0_-\rangle$, such as depicted in FIG.1 in the case of $QED$, or can they be obtained by relying on calculations of $S$-matrix elements like the Casimir effects?
\par
Condensates being recognised physical realities in the $QFT$ standard model context \cite{Jaffe},  if they can exclusively be calculated out of the vacuum energy density related to $\langle 0_+|0_-\rangle$, then this vacuum energy density could be regared as endowed with some physical reality also. This is the stake of the question.

A famous and most important example is that of the \emph{dynamical  chiral symmetry breaking mechanism in $QCD$ or $QED$}. A measure of this chiral symmetry breaking phenomenon is provided by the Lorentz scalar and gauge invariant fermionic condensate $\langle\bar{ \Psi}\Psi(x)\rangle$, and is therefore recognized as an \emph{order parameter} of this symmetry. To simplify somewhat while preserving the point, in an abelian theory like $QED$ this order parameter can be obtained  by calculating the condensate according to the relation,
\begin{equation}\label{une}
\langle\bar{ \Psi}\Psi(x)\rangle=- \frac{1}{V} \frac {\partial}{\partial m}\,\ln \,\langle 0_+|0_-\rangle\end{equation}where $m$ stands for a relevant fermionic mass and $V$ for the overall spacetime volume. In this way, explicit reference is made to the series of vacuum diagrams of FIG.1, related to the would be vacuum energy density. In Ref.\cite{Herbtg} for instance, this equation was used successfully to evaluate the fermionic condensate in the massive Schwinger model (\textit{i.e.}, massive $QED$ at two spacetime dimensions, \emph{non-integrable} contrary to the massless version). But using (\ref{une}) is in no way mandatory, as the same result could be achieved relying on a $4$-point \emph{Green's function} calculation \cite{fghf}, parts of which related to some scattering processes among particles.\par
 More recently the same procedure has been successfully extended to $QCD$ \cite{tgpt} avoiding to create a link of necessity between the condensate and a would be non-zero vacuum energy density, as was questioned in \cite{Jaffe}.  As well known \emph{connected} Green's function don't even `see' the \emph{pure phase} factor $\langle 0_+|0_-\rangle$. 
 \par
 After all, that particles involved in selected configurations of scattering processes may `feel' the presence of condensates at the energy scales where they come about seems quite plausible. 
 \par
 Discussing this point further falls beyond the scope of the present paper but it is worth signalling also that in the \emph{light front formulation} of $QCD$, dynamical  chiral symmetry breaking \emph{is not} a property of the hadron-less vacuum state of $QCD$ \cite{Casher}. The questioning  in the conclusion of \cite{Jaffe} could very well find a clue in that, as suspected by P. Carruthers \emph {.. it is a mistake to identify the ground state with the vacuum} \cite{Carruthers}. The condensate calculation in \cite{tgpt} can be seen as an illustration of this wisdom.

\end{document}